\documentclass[twocolumn, usenatbib]{mnras}
\usepackage[utf8]{inputenc}
\usepackage{amssymb}
\usepackage{graphicx}
\usepackage{amsmath}
\usepackage{float}
\usepackage{subfigure}
\usepackage{hyperref}

\begin{document}

\title[Quantifying Excess Power from RFI in EoR Measurements]{Quantifying Excess Power from Radio Frequency Interference in Epoch of Reionization Measurements}
\author[M. J. Wilensky et al.]{Michael J. Wilensky,$^{1}$\thanks{mjw768@uw.edu} Nichole Barry,$^{2, 3}$ Miguel F. Morales,$^{1, 4}$ \newauthor{Bryna J. Hazelton,$^{1, 5}$Ruby Byrne$^{1}$}
\\
$^{1}$Department of Physics, University of Washington, Seattle, WA 98195, USA\\
$^{2}$School of Physics, The University of Melbourne, Parkville, VIC 3010, Australia\\
$^{3}$ARC Centre of Excellence for All Sky Astrophysics in 3 Dimensions (ASTRO 3D), Australia\\
$^{4}$Dark Universe Science Center, University of Washington, Seattle, 98195, USA\\
$^{5}$eScience Institute, University of Washington, Seattle, WA 98195, USA}

\date{Submitted to Monthly Notices of the Royal Astronomical Society April 15 2020, accepted on August 06 2020.}

\maketitle

\begin{abstract}
    We quantify the effect of radio frequency interference (RFI) on measurements of the 21-cm power spectrum during the Epoch of Reionization (EoR). Specifically, we investigate how the frequency structure of RFI source emission generates contamination in higher-order wave modes that is much more problematic than smooth-spectrum foreground sources. Using a relatively optimistic EoR model, we find that even a single relatively dim RFI source can overwhelm the EoR power spectrum signal of $\sim10\text{ mK}^2$ for modes $0.1 \text{ }h\text{ Mpc}^{-1} < k < 2 \text{ }h\text{ Mpc}^{-1}$. If total apparent RFI flux density in the final power spectrum integration is kept below 1 mJy, an EoR signal resembling this optimistic model should be detectable for modes $k < 0.9\text{ }h\text{ Mpc}^{-1}$, given no other systematic contaminants and an error tolerance as high as 10\%. More pessimistic models will be more restrictive.  These results emphasize the need for highly effective RFI mitigation strategies for telescopes used to search for the EoR.
\end{abstract}


\section{Introduction}


The Epoch of Reionization (EoR) is a cosmological period in which the content of the universe transitioned from being mostly neutral to being mostly ionized, as we see it today. Understanding the EoR will inform other areas of cosmology such as structure formation and the expansion history of the universe. For reviews on the study of the EoR, see \citet{Furl2006}, \citet{Morales2010}, and \citet{Liu2019}. One way to probe the EoR is to measure the power spectrum of fluctuations in the brightness temperature of 21-cm radiation emitted from the neutral Hydrogen hyperfine transition. These types of measurements are performed using radio interferometers such as the Murchison Widefield Array (MWA) \citep{Tingay2013, Wayth2018}, the Hydrogen Epoch of Reionization Array (HERA) \citep{DeBoer2017}, the LOw Frequency ARray (LOFAR) \citep{LOFAR2013}, the Precision Array for Probing the Epoch of Reionization (PAPER) \citep{Parsons2010}, and the Giant Metrewave Radio Telescope (GMRT) \citep{GMRT}, as well as the developing Square Kilometer Array \citep{SKA}. Constructed for a given redshift, the 21-cm power spectrum tells us about the scales of neutral hydrogen at that moment in cosmological history.  

 A common theme that emerges from attempted measurements of the EoR power spectrum is that systematics, particularly those that yield frequency-dependent effects, must be understood with a high dynamic range in order for an EoR detection to be feasible. These systematics come in many forms. For instance, bright radio emission from astrophysical foregrounds is typically 4-5 orders of magnitude brighter than the expected reionization signal. These foregrounds dominate the EoR signal in a region of power spectrum space known as the foreground wedge \citep{Datta2010, Morales2012, Trott2012, Parsons2012, Thyagarajan_2013}. Moreover, errors in calibration that arise from incomplete knowledge of these sources can themselves provide contamination of power spectra, manifest in calibration errors \citep{Datta2010, Barry2016, Trott2016, Patil2016, Patil2017, Ewall-Wice2017, Li2018, Li2019, Byrne2019, Joseph2020, Kern2020b}. Other effects involve subtle instrumental behavior that would be negligible in other contexts, including, but not limited to, effects arising from baseline layout \citep{Hazelton2013, Murray2018}, complexities involving the primary beam \citep{Beardsley2016, Ewall-Wice2016,Thyagarajan2016, Barry2019b, Fagnoni2019, Joseph2020}, and internal signal chain reflections \citep{Beardsley2016, Barry2019b, Kern_2019, Kern2020a, Kern2020b, Li2019}, . Another important class of contaminants, which is the focus of this work, are radio signals such as FM radio broadcasts and other anthropogenic transmissions that are referred to in radio astronomy as radio frequency interference (RFI). 

Reception of RFI poses problems for many radio astronomy applications. RFI can often be exceptionally bright compared to distant astrophysical radio sources, and there are many different mechanisms by which RFI can enter an array. Such mechanisms include self-generated RFI within a telescope, direct reception of RFI from a nearby transmitter, long range reception of RFI due to tropospheric ducting \citep{Sokolowski2016}, broadband emission from lightning \citep{Sokolowski2016}, direct emission from satellites such as ORBCOMM \citep{Sokolowski2016}, as well as reflections off of aircraft \citep{Wilensky2019}, the moon \citep{McKinley2013, McKinley2018}, meteors \citep{Zhang2018}, and satellites \citep{Zhang2018}. The prolific means by which RFI can contaminate radio data make it a ubiquitous problem, and so there has been much study devoted to mitigation methods. While an exhaustive review of general RFI mitigation techniques employed in different radio contexts would be outside the scope of this work, we note a few post-correlation flagging and filtering techniques that have been employed on EoR data so far, namely, \textsc{AOFlagger} \citep{Offringa2012, Offringa2015}, deep learning techniques with convolutional neural networks \citep{Kerrigan2019}, the Watershed RFI algorithm \citep{Kerrigan2019, Roerdink}, as well as \textsc{SSINS} \citep{Wilensky2019}. 

While there is a strong consensus that RFI flagging is a highly important step in pre-processing EoR data \citep{Barry2019b, Li2019, Mertens2020}, and much effort has been devoted to removing RFI contaminated data from EoR studies, there has been relatively little study to specifically characterize the manner in which undetected RFI systematically affects EoR detection efforts. Interestingly, RFI excision itself can induce chromatic structure in radio data that can systematically bias power spectrum estimates, which is specifically studied in \citet{Offringa2019a}. In general, any given mitigation method is bound to leave undetected RFI at some level. For instance, \citet{Offringa2013} presents a statistical estimate for the apparent brightness of undetected RFI in LOFAR data after \textsc{AOFlagger}, along with an empirical limit on the contamination levels judged from imaging RFI-flagged data. Another example is \citet{Barry2019b}, in which two rounds of RFI flagging were performed on MWA EoR data: first with \textsc{AOFlagger}, and then with \textsc{SSINS}. In this study, \textsc{SSINS} found a substantial number of faint digital television (DTV) interference events remaining after the first round of flagging. Furthermore, removal of any 2-minute snapshots containing any DTV remnants helped improve EoR limits, where the improvement was quite substantial in a certain subset of the data. A natural line of inquiry that follows from this is to investigate the level to which undetected RFI occludes EoR detection. The purpose of this work is to provide a quantitative, theoretical expectation for the systematic bias in the power spectrum that results from undetected RFI, and to thereby quantify a tolerance for RFI contamination in EoR data based on the expected brightness of the EoR signal. 

In section \S\ref{sec:formal_intro}, we lay the foundation for the theoretical formalism used in this paper and state assumptions that are used. In \S\ref{sec:results}, we first calculate the expected power spectrum of several common types of RFI source using this formalism. Using a fully functional imaging and power spectrum pipeline, we then simulate RFI power spectra and compare them to simulated power spectra of sources from the GLEAM sky catalog \citep{Hurley-Walker2017} as well as theoretical expectations for the EoR signal. In \S\ref{sec:conc}, we draw conclusions from this work.

\section{Theoretical Formalism}
\label{sec:formal_intro}

In this section we establish the theoretical formalism that is used for the power spectrum calculations in this work. This includes choices of notation, definitions of useful quantities, and assumptions about quantities that affect the calculation.




\subsection{The Power Spectrum Estimator}

Since the goal of this work is to extract the effect that an RFI source has on the power spectrum, we will make several simplifying assumptions in order to ease theoretical calculations. We will later compare these theoretical calculations to realistic in-situ simulations wherein qualities of the instrumental measurement process are included, so that we can be sure that the gross effects predicted from the theoretical calculations are still present along with the nuances that arise from measurement. For analytic descriptions of EoR power spectra that include effects of the instrument and analysis choices, see \citet{Liu2014a, Liu2014b}.

Astrophysical foreground contamination is largely contained within a wedge-shaped region of the cylindrically averaged Fourier domain known as the foreground wedge, e.g. \citet{Datta2010, Morales2012, Trott2012, Parsons2012, Thyagarajan_2013}. Without extremely accurate foreground removal, this contamination excludes lower-order line-of-sight modes in power spectrum measurements. The depth of exclusion depends linearly on the length of the perpendicular mode in question, hence the wedge shape. The higher-order line-of-sight modes with strongly diminished foreground contamination form the EoR window, e.g. \citet{Liu2014a, Liu2014b}. As is shown in \S\ref{sec:results}, the most problematic feature of RFI seen in this analysis is that it provides very strong excess power in the EoR window. This contribution arises strictly from the intrinsic frequency structure of RFI. With this in mind, we will ignore the chromatic point-spread function that leads to the foreground wedge in order to simplify the theoretical calculations. Moreover, concentrating on the spectral characteristics of the RFI makes the theoretical derivation equally applicable for all EoR power spectrum analyses \citep{Morales2019}.

We also choose to ignore the fine details of the bandpass response of the instrument, chromaticity of the primary beam, as well as chromatic errors in calibration. These effects are all deeply important in power spectrum estimation \citep{Beardsley2016, Barry2016, Byrne2019, Ewall-Wice2017, Patil2016}. While generally not negligible, they are beyond the scope of the theoretical calculations in this work. We will include the fact that an instrument possesses a limited field of view and is only sensitive to some range of radiation frequencies. Since it is highly relevant to this analysis and convenient to implement theoretically, we will also include a frequency tapering function for this range of frequencies. A frequency tapering function is commonly used in power spectrum estimation that gracefully limit foreground spillover into the EoR window \citep{Thyagarajan_2013} and ensure that a sufficiently narrow range of frequencies is used to prevent the influence of cosmological evolution \citep{Morales2004}. 

Bearing these assumptions in mind, we define the power spectrum estimator using the following expression:
\begin{equation}
    P(\bold{k}) = \frac{1}{V_{\mathcal{M}}}\bigg|\int_{\mathcal{M}} \text{d}^3\bold{r}A\big(\boldsymbol{\theta}(\bold{r})\big)\Psi\big(f(r_\parallel)\big)I\big(\boldsymbol{\theta}(\bold{r}), f(r_\parallel)\big)\text{e}^{-\text{i}\bold{k}\cdot\bold{r}}\bigg|^2,
    \label{eq:ps_def}
\end{equation}
where \(\mathcal{M}\) is the cosmological region of space for which the power spectrum is being constructed, \(V_\mathcal{M}\) is the cosmological volume of that region, \(\bold{r} = (r_x, r_y, r_\parallel)\) is a position vector in comoving cosmological co-ordinates, and $\bold{k} = (k_x, k_y, k_\parallel)$ is the Fourier dual to that vector. In this analysis, the primary beam, $A$, the frequency tapering function, $\Psi$, and the brightness temperature map, $I$, are more readily expressed and understood in terms of angular co-ordinates $\boldsymbol{\theta} = (\theta_x, \theta_y)$ and the observed frequency of emission, $f$, hence the listed arguments of these functions above. For the sake of notational brevity, we will also occasionally make use of the vectors $\bold{k_\perp} = (k_x, k_y)$, $\bold{r_\perp}=(r_x, r_y)$, $\bold{u} = (u, v)$. Conversions between different co-ordinate systems and estimators can be found in various references, such as \citet{Morales2004} and \citet{Liu2014a}. For the relatively narrow frequency bands considered in typical EoR analyses, the relation between comoving line-of-sight distance to a narrow emitter and its observed frequency of emission is very nearly linear i.e. for the purposes of calculation, the following relationship is justified:
\begin{equation}
    r_\parallel(f) \approx \beta f + c_0.
    \label{eq:lin_approx}
\end{equation}
The slope of this linear relation, $\beta$, is given in terms of the speed of light, the central redshift of the observation, the frequency of the 21-cm radiation in the rest frame, and other cosmological parameters. See equation 4 of \citet{Morales2004}. Physically, $c_0$ is the extrapolated comoving distance at infinite redshift, which is not meaningful by itself since the relationship is not genuinely linear. In this case, the constant merely serves to make the linear relation consistent with the universe's geometry for the relevant redshift. A linear relationship allows the Fourier transform over the line of sight coordinate to be directly written as a Fourier transform over frequency instead, which is both helpful when considering observational quantities and also what is done in the practice of power spectrum estimation.
\subsection{Calculating a Power Spectrum for Point Sources}

First, we calculate the power spectrum for a single point source. For this case, equation \ref{eq:ps_def} can be written as 
\begin{equation}
    \begin{split}
        P(\bold{k}) = \frac{1}{V_\mathcal{M}}\bigg|\beta\int_{f_L}^{f_U}\text{d}f\text{e}^{-\text{i}k_\parallel r_\parallel(f)}\Psi(f)D_M(f)^2\phi(f)\kappa(f)\\
        \times\int_{FoV}\text{d}^2\boldsymbol{\theta}\text{e}^{-\text{i}D_M(f)\bold{k_\perp}\cdot\boldsymbol{\theta}}I_0\delta^2(\boldsymbol{\theta}-\boldsymbol{\theta_0})A(\boldsymbol{\theta})\bigg|^2,
    \end{split}
\end{equation}
where we have chosen to write the transverse spatial integrals as angular integrals since that is a more natural co-ordinate system in which to express the flux of a point-source as well as the frequency tapering function and primary beam (whose achromaticty is now explicitly manifested). Here, $f_L$ and $f_U$ represent the lower and upper bounds of the frequencies observed by the instrument, $D_M(f)$ is the transverse comoving distance \citep{Hogg1999} introduced in the co-ordinate conversion from $\bold{r_\perp}$ to $\boldsymbol{\theta}$, $I_0\phi(f)$ is the flux density of the source measured in Jy ($I_0$ is its total flux), $\kappa(f)$ converts this flux density into brightness temperature, and the Dirac delta function represents the point-like nature of the source at angular position $\boldsymbol{\theta_0}$. Evaluating the angular integral and assuming a flat cosmology\footnote{This lets one equate the transverse and line-of-sight comoving distances. See \citet{Hogg1999}.} leaves
\begin{equation}
    \begin{split}
        P(\bold{k}) = \frac{1}{V_\mathcal{M}}\bigg|\beta\int_{f_L}^{f_U}\text{d}f\text{e}^{-\text{i}(k_\parallel + \bold{k_\perp}\cdot\boldsymbol{\theta_0})r_\parallel(f)}\\
        \times I_0^{\text{app}}\phi(f)\kappa(f)\Psi(f)r_\parallel(f)^2\bigg|^2,
    \end{split}
    \label{eq:PS_num}
\end{equation}
where $I_0^{\text{app}}=A(\theta_0)I_0$ is the apparent flux of the source. At this point, even with the simplification afforded by equation \ref{eq:lin_approx}, this integral is only analytically calculable for special cases. To go further will require a specification of the frequency dependence of the source and typically some numerical methods. 

Placing the source at phase tracking center eliminates the \(\bold{k_\perp}\) dependence in the power spectrum, thus the cylindrically averaged power spectrum of a source at phase tracking center will be constant along lines of constant $k_\parallel$. Simulation results do not show a strong difference between power spectra of centered and off-center RFI sources, particularly within the EoR window. For this reason, we will only show theoretical power spectra for sources at phase center.  

\subsection{Imposing a Frequency Dependence}

In general, RFI sources may take on a great variety of frequency profiles. Many transmissions only occupy a single channel in an EoR analysis, although some can be relatively wide compared to the observing band.   

Two types of common and relatively broad RFI signals observed in EoR analyses include DTV and digital audio broadcasting (DAB). These types of signals are well-approximated by top-hats in frequency over their allocation.\footnote{Examples of technical standards for such signals can be searched for at \url{http://www.etsi.org/standards-search}. DTV broadcasts may be referred to as digital video broadcasting (DVB) in such manuals.} The widths and locations of these profiles are usually subject to matters of protocol in their country of origin. For this work we use the Western Australian 7 MHz wide allocations, as is frequently observed by the MWA. For example, see \citet{Offringa2015}, \citet{Sokolowski2016}, and \citet{Wilensky2019}. 

We would write a top-hat dependence of width \(\Delta f\) and center frequency \(f_0\) like so:\footnote{The normalization in equation \ref{eq:top-hat} ensures consistent units for $I_0$ in equation \ref{eq:PS_num}.}
\begin{equation}
    \phi(f) = \frac{1}{\Delta f}\Pi\bigg(\frac{f-f_0}{ \Delta f}\bigg),
    \label{eq:top-hat}
\end{equation}
where $\Pi(x)$ is the rectangle function, given by
\begin{equation}
    \Pi(x) = \begin{cases}
                1, & \text{if}\ |x| < \frac{1}{2} \\
                  0, & \text{otherwise}.
    \end{cases}
\end{equation}
Given the slight complication in frequency dependence from converting to brightness temperature and applying a tapering function, we choose to evaluate the integral in equation \ref{eq:PS_num} numerically for the comparisons made in this work.

A sufficiently narrow signal (the resolution of the analysis or finer) can be formally handled using a Dirac delta function frequency dependence. This renders the Fourier transform over frequency analytically straight-forward. The result is a constant power on all k-modes.

\begin{figure*}
    \centering
    \includegraphics[scale=0.45]{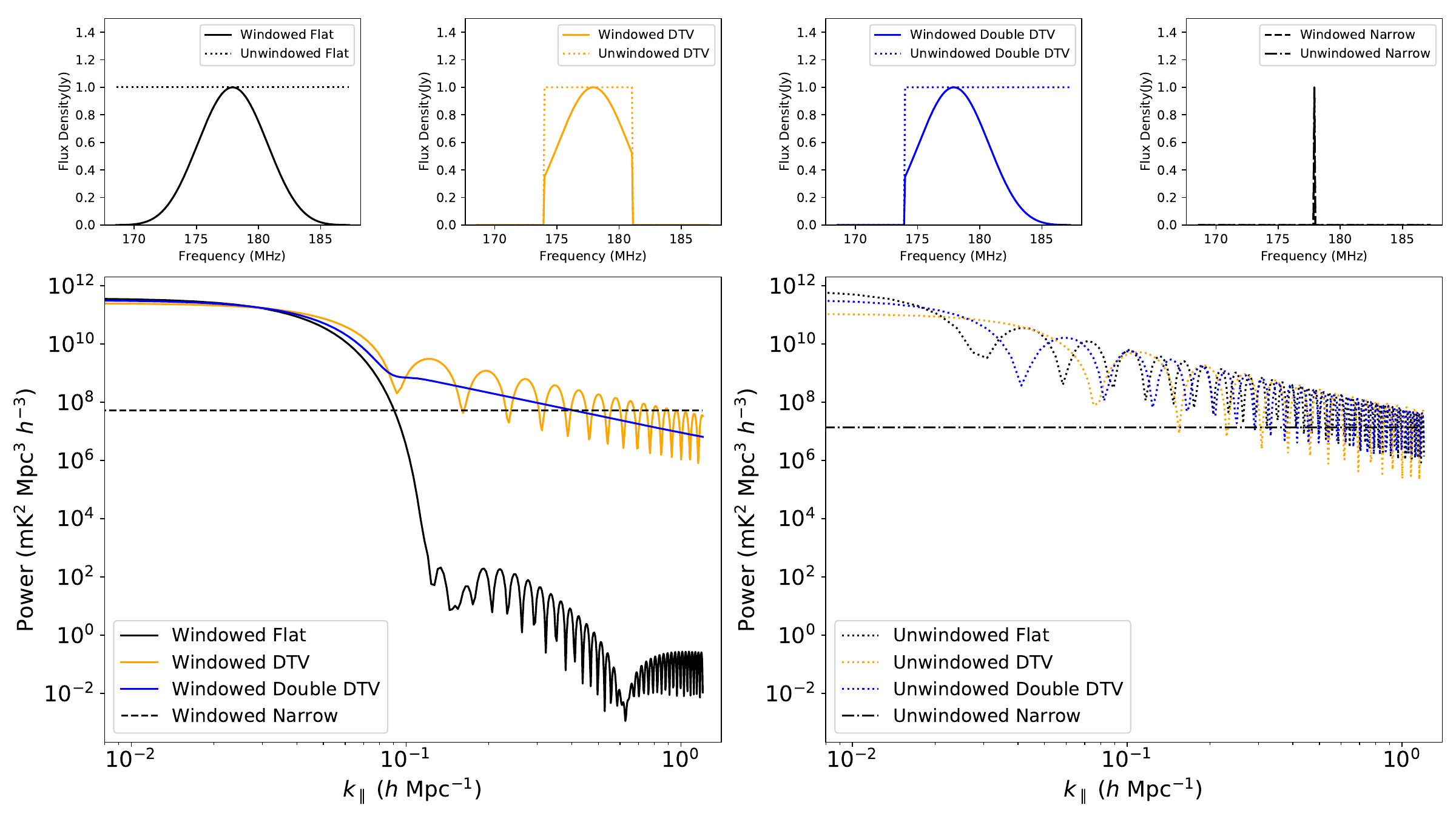}
    \caption{Theoretical frequency structures and power spectra along the line of sight for various 1-Jy point-sources. The top panels show the windowed (solid, narowband is dashed) and unwindowed (dotted, narrowband is dashed-dotted) frequency structure of each source in Jy, while the bottom panels show the corresponding power spectra of the windowed (bottom left) and unwindowed (bottom right) sources. The sources are placed at the  phase center of a highly idealized instrument detailed in \S\ref{sec:formal_intro}, so there is no $\bold{k_\perp}$ dependence. For the broader sources, the unwindowed spectra are essentially identical to one another except in the lowest-ordered $k$-modes. In the windowed case, we see that the flat-spectrum (foreground) source contamination has a sharp falloff starting at modes greater than $\sim\text{0.05 h Mpc}^{-1}$. This is exactly the desired effect of the Blackman-Harris window function. However, the window function does not have nearly the same effect on the RFI sources, whose contamination remains many orders of magnitude greater than that of the foreground source. The narrowband source is in the middle of the observing band, so the window function has no effect on the frequency structure. However, the power spectra differ because the effective cosmological volume probed in the frequency-tapered case is smaller than in the non-tapered case.}
    \label{fig:freq_to_power_BH}
\end{figure*}

Due to the relative ease of evaluating power spectra for narrowband point sources, one may glean information about ensembles of narrowband point-sources using this analytic machinery. We relegate the precise mathematical details to Appendix \ref{sec:app} and summarize the results here. In principle, any two sources could destructively interfere for some mode in the power spectrum, however, any two sources will interfere constructively for modes perpendicular to the separation vector of the sources. The power of any ensemble for some given mode is bounded from above by the case of total constructive interference; the power is proportional to the square of the sum of the fluxes. The modes for which this can be achieved are intimately related to the angular distribution of the emitters. For a random ensemble drawn from a uniform distribution over the sky and whose true flux distribution is independent of its uniform angular distribution, the expected power (average over realizations) is bounded from below by an incoherent sum of powers; it is proportional to the quadrature sum of the fluxes i.e. a linear sum of individual powers. For this particular distribution, baselines longer than the inverse width of the primary beam should observe relatively little coherence. Since this is the minimum length of a physical baseline, an angularly uniform distribution of RFI emitters are expected to add power incoherently for any mode in an estimated power spectrum, although specific realizations can defy this expectation. In conclusion, it is expected that adding many uniformly distributed RFI sources should tend to increase contamination relative to a single source in a predictable way. Therefore, if we can understand the strength of a single RFI source on a power spectrum measurement relative to the EoR, we can quantify the degree to which RFI needs to be mitigated to make an EoR detection feasible.

\section{Results}
\label{sec:results}

In this section, we apply the theoretical formalism in \S\ref{sec:formal_intro} to make rough predictions about the general behavior and contamination levels of RFI in the power spectrum. In order to compare these predictions to realistic power spectra, we also simulate power spectra of RFI sources and sources from the GLEAM sky catalog using the \textsc{FHD}\footnote{https://github.com/EoRImaging/FHD}/\textsc{$\varepsilon$ppsilon}\footnote{https://github.com/EoRImaging/eppsilon} power spectrum pipeline \citep{Sullivan2012, Barry2019a}. Ultimately we find that, in both simulation and theory, RFI provides orders-of-magnitude more contamination than typical point-like foreground sources within the EoR window. We do not estimate power spectra for diffuse foregrounds, which are substantial. Including these may make foregrounds more competitive with RFI on some modes.

\subsection{Theoretical Results for RFI Power Spectra}

We specifically consider power spectra from sources with the following frequency dependences: flat broadband, a single DTV channel, two simultaneous frequency-adjacent DTV channels (both received from the same position e.g. from an aircraft reflection), and a narrowband source. A flat, broadband source emulates an astrophysical foreground source so that there is a baseline contamination level to which we can compare that afforded by band-limited RFI. For all sources, we numerically evaluate the integral in equation \ref{eq:PS_num}, with and without a Blackman-Harris tapering function. The results for a collection of 1 Jy sources at phase tracking center are shown in Figure \ref{fig:freq_to_power_BH}. We used the same observing band as was used in the final limit in \citet{Barry2019b}, centered on redshift 7 for the 21-cm line. Of striking significance is that the RFI sources universally provide dramatic contamination in the higher order line-of-sight modes in the power spectrum compared to the flat-spectrum (foreground) source due to sharp spectral cutoffs within the observing band. 

\begin{figure*}
    \centering
    \begin{subfigure}[]{
        \includegraphics[width=0.3\linewidth]{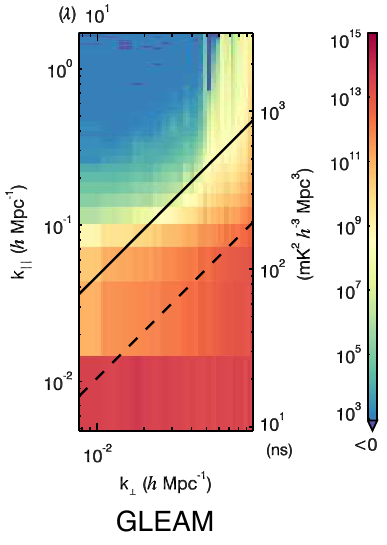}}
    \end{subfigure}
    \begin{subfigure}[]{
        \includegraphics[width=0.31\linewidth]{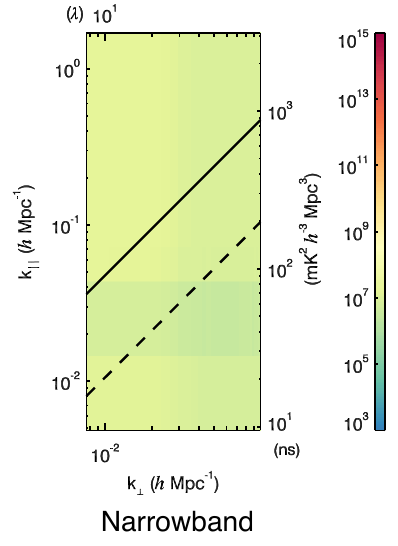}}
    \end{subfigure}
    \begin{subfigure}[]{
        \includegraphics[width=0.298\linewidth]{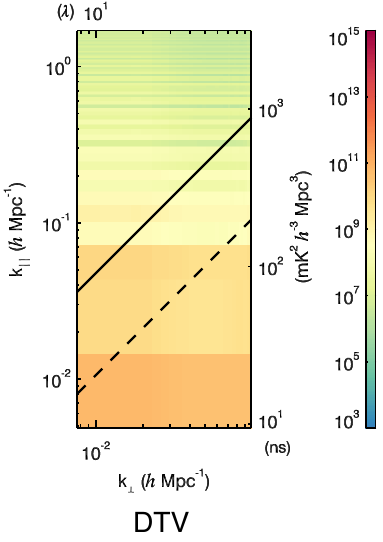}}
    \end{subfigure}
    \caption{Cylindrical power spectra for simulations with GLEAM (left), narrowband (middle), and DTV (right) sources. The solid line marks the scales corresponding to the horizon, while the dashed line demarcates the scale of the primary field of view. Smooth-spectrum source contamination is predominantly contained within the foreground wedge, which extends everywhere below the solid line. Above the solid line is the EoR window. The structure seen in these plots generally match our theoretical expectations. The EoR window contamination of a single 1-Jy RFI source is substantially higher than what is included from GLEAM.}
    \label{fig:2d_sim_PS}
\end{figure*}

As mentioned in \S\ref{sec:formal_intro}, a frequency tapering function is typically employed to reduce power leakage of foregrounds into the EoR window, thus boosting the dynamic range of the measurement. The desired effect is shown in Figure \ref{fig:freq_to_power_BH}, where the flat-spectrum source, which resembles an astrophysical foreground, displays a sharp falloff at higher $k_\parallel$ relative to its unwindowed power. This effect occurs because the tapering function smoothly approaches zero at the boundaries of the observing band, eliminating sharp spectral cutoffs for this source \citep{Harris1978}.

On the other hand, the RFI sources are band-limited, and so the tapering function fails to eliminate the sharp spectral cutoffs of the RFI sources. Comparing the power spectra of windowed and unwindowed RFI sources, we observe that the tapering function has little effect on overall contamination levels, in great contrast to the flat-spectrum source. Generically, as the bandwidth of a source approaches the bandwidth of the observing window, and thus more closely resembles a foreground source, the EoR window contamination is more effectively reduced by the frequency tapering function. However, we see that even in the case of the two-channel DTV event, which occupies nearly three-quarters of the band, the window function makes hardly any difference relative to the single-channel DTV event. Additional calculations not shown indicate that an emitter must occupy greater than 95\% of the observing band to appreciably close the several order-of-magnitude gap between the RFI and foreground contamination seen in Figure \ref{fig:freq_to_power_BH}. Some experimentation shows that different tapering functions may close this gap more effectively, however this typically comes at the expense of the effective and crucial dynamic range boost offered by the Blackman-Harris window.

This suggests an interesting and very simple mitigation strategy for appropriately sized sources such as DTV RFI. If the observing band is set equal to the band allocated for the RFI source, then the tapering function will affect the RFI power spectrum in the same way the flat-spectrum source was affected in Figure \ref{fig:freq_to_power_BH}. Of course, any narrower RFI source within the same allocation will be relatively unaffected by this choice. While regional radio allocations are set so that particular types of signals are broadcast over certain frequencies, RFI other than the regionally allocated signals can be observed through several mechanisms e.g. intermodulation products arising from nonlinearities in the instrument signal path. Furthermore, the size of the observing band is subject to important constraints, including sensitivity requirements for EoR detection and the need to preserve the assumption of isotropy over the probed cosmological volume. Therefore, this particular strategy is extremely limited in scope, and should not be the primary mitigation strategy. The enormous contamination of RFI in the EoR window suggests that specialized post-excision mitigation strategies are needed if significant RFI sources are missed by excision algorithms. We specifically discuss the degree to which RFI sources need be mitigated in terms of allowable apparent flux density in \S\ref{sec:budget}.

The effectiveness of the tapering function on a narrowband source is strictly a function of the location of that source within the observing band. The window function multiplicatively adjusts the constant power level provided by the source by the square of the value of the window function at the frequency of the source. Here, we have chosen to keep the narrowband source at the center of the window, so that we could see the full effect of an unmitigated narrowband source.
 
There is also an interesting difference between the two-channel DTV event and the single-channel DTV event, which is that the single-channel event exhibits a lobed structure as a function of $k_\parallel$, while the two-channel event does not. This is an interaction between the lobes of the Fourier transforms of the DTV signal and tapering function that depends on the location of the event within the observing band \citep{Harris1978}. This effect is of little practical consequence, since the overall contamination levels are roughly identical regardless of the presence of lobes. 

As mentioned in \S\ref{sec:formal_intro}, these theoretical calculations ignore many important aspects of measuring EoR power spectra with radio interferometers. In order to check these predictions against realistic power spectra made from instrumental visibilities and a fully functioning analysis pipeline, we employed the simulation capabilities of the \textsc{FHD/$\varepsilon$ppsilon} pipeline. We also simulate the power spectrum of sources from the GLEAM catalog, so that we can compare simulated RFI power spectra to a power spectrum of a sky full of foregrounds.

\subsection{Simulated RFI Power Spectra}
\label{sec:simulation}

 We show cylindrical power spectra for sources from the GLEAM catalog and two different RFI sources in Figure \ref{fig:2d_sim_PS}. In these simulations, we used the MWA Phase I \citep{Tingay2013} as our example instrument, thus capturing a realistic baseline distribution and beam pattern for an interformeter. The beam is simulated using an average embedded element and mutual coupling model \citep{Sutinjo2015}, chosen at the central frequency of observation (about 178 MHz). This allows us to accurately depict some typical chromatic effects of an instrument and analysis pipeline. To simulate each RFI source, we modeled a single fictitious, phase-centered, flat-spectrum, 1 Jy source, and simulated visibilities using \textsc{FHD}. Using \textsc{pyuvdata}\footnote{https://github.com/RadioAstronomySoftwareGroup/pyuvdata} \citep{Hazelton2017}, we then modified the frequency structure of the fictitious source to match our desired RFI sources. We then passed these visibilities back through \textsc{FHD} to create the input HEALPix cubes \citep{Gorski2005} for \textsc{$\varepsilon$ppsilon}. The simulated GLEAM sources lie in the same observing field as was used in \citet{Barry2019b}, which is a region of the sky with minimal bright sources and low sky temperature. 
 As expected from our theoretical work, the narrowband source has a nearly constant power spectrum, while the DTV source has a lobed structure along $k_\parallel$ and relatively little structure along $k_\perp$. The power spectrum of the simulated GLEAM catalog demonstrates how foreground power is largely confined to the foreground wedge, with very little power escaping into the EoR window. Because the RFI sources are at phase center, they do not exhibit any sort of foreground wedge. While an off-center RFI source would not be immune to the effects that cause the foreground wedge, the amount of power leakage into higher line-of-sight modes from the wedge (a $\sim 0.01- 0.1\%$ effect in the foregrounds of Figure \ref{fig:2d_sim_PS}(a)) is subdominant to the spillover that occurs from the band-limited nature of the RFI (a $\sim 10\%$ effect for the DTV power spectrum in Figure \ref{fig:2d_sim_PS}(c)). We can see that this spillover drastically contaminates the EoR window.

\begin{figure*}
    \centering
    \includegraphics[scale=0.5]{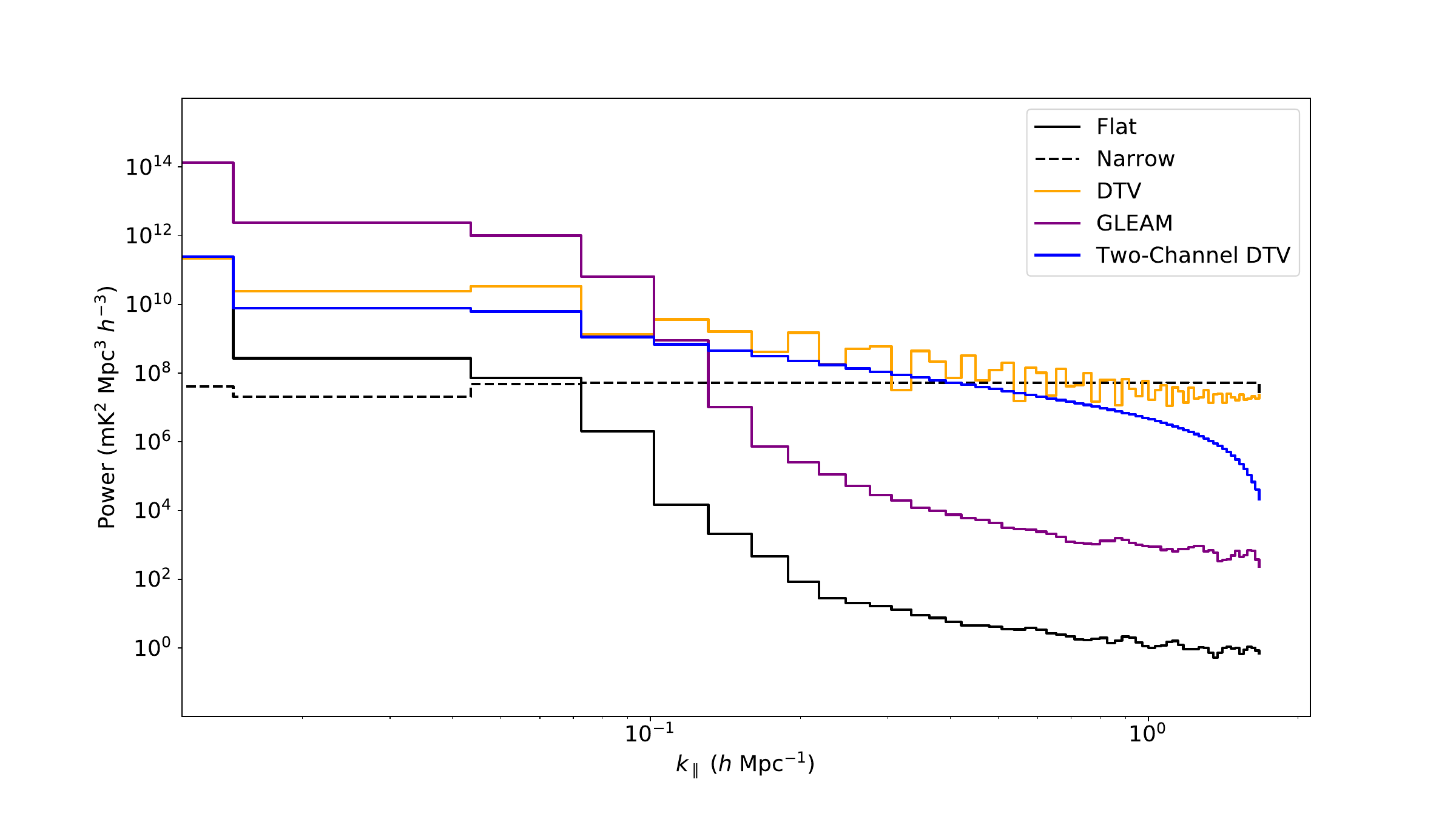}
    \caption{Simulated power spectra (averaged over $\bold{k}_\perp$) analogous to the theoretical spectra in Figure \ref{fig:freq_to_power_BH}, with an additional line for a simulated power spectrum of sources from the GLEAM catalog. The RFI spectra show a strong correspondence with the theoretical prediction. The two-channel DTV power spectrum is seemingly smooth compared to the lobed single-channel power spectrum, although the difference in power is more pronounced at higher modes than theoretically predicted. The falloff for the 1-Jy foreground source is similar, although the theoretical lobed structure is not visible. There is still a large difference in contamination between the foreground source and the RFI sources. Moreover, even the contamination from the simulated GLEAM catalog has several orders of magnitude less contamination than a single 1-Jy DTV or narrowband source at high $k_\parallel$.}
    \label{fig:1d_sim}
\end{figure*}

The cylindrical power spectra from Figure \ref{fig:2d_sim_PS}, averaged over the $k_\perp$ axis, are shown in Figure \ref{fig:1d_sim}, along with some additional simulated power spectra for comparison to Figure \ref{fig:freq_to_power_BH}. We can see that in a realistically simulated power spectrum, there is still a several order of magnitude discrepancy between the contamination of astrophysical foregrounds and that of RFI within the EoR window, thus confirming the earlier theoretical predictions. A natural line of investigation is to see what level of RFI is sustainable given the high dynamic range required to measure the EoR signal in the power spectrum.

\subsection{How to Develop an RFI Budget for EoR Detection}
\label{sec:budget}

The final RFI budget for a given analysis depends on several qualities of the experiment. The type of RFI signal observed by the instrument as well as details about the RFI environment of the telescope, such as the number of emitters and their angular distribution, all affect the amount of RFI contamination. Furthermore, the cosmological modes of interest and actual strength of the EoR signal in those modes determines the amount of allowable contamination. To design a budget, the analyst must also decide on a model for the EoR. The final RFI budget designed by the analyst will of course be affected by the choice of model. An RFI power spectrum that is beneath the expected EoR signal by a significant amount for the modes of interest gives a budget for those modes. 

In this work, we design some example budgets for different RFI realizations, adopting the fiducial model used in \citet{Barry2019b}. This is a relatively optimistic model made using \textsc{21cmFAST} \citep{Mesinger2011} and an astrophysical parameterization from \citet{Park2019}. We focus on modes between 0.1 and 2 $h\text{ Mpc}^{-1}$, which represents the foreground avoidance strategy employed in MWA EoR analyses. For these modes the model predicts a signal strength of order $\Delta^2\sim10\text{ mK}^2$, where $\Delta^2=(k^3 / 2\pi^2)P(k)$. A foreground removal strategy might endeavour to measure the EoR on lower modes where some models expect a stronger EoR signal, which would affect the overall RFI budget compared to these examples.

\begin{figure*}
    \centering
    \begin{subfigure}[]{
        \includegraphics[width=0.48\linewidth]{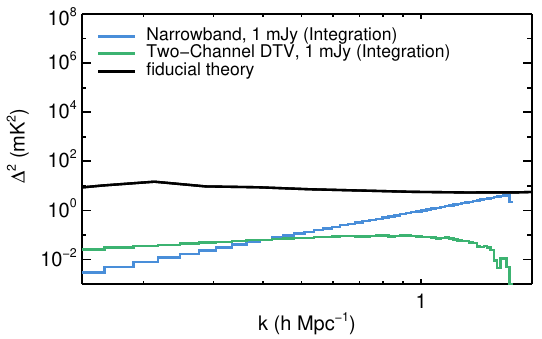}}
    \end{subfigure}
    \begin{subfigure}[]{
        \includegraphics[width=0.48\linewidth]{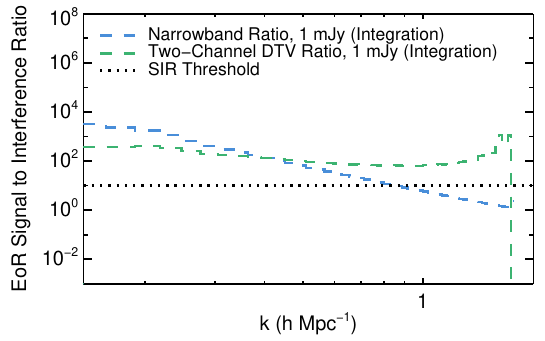}}
    \end{subfigure}
    \caption{(a) 1 mJy narrowband (blue) and two-channel DTV (green) RFI power spectra, along with the fiducial EoR model from \citet{Barry2019b} (black). We can see that even at an integrated flux density of 1 mJy, a narrowband RFI source can compete with the EoR signal at higher $k$-modes. The DTV source has flatter scaling as a function of $k$, and so is substantially lower then the EoR model on all modes shown. (b) The EoR signal to RFI power ratios (dashed) for the spectra shown in (a), along with a level at 10, indicating the minimum allowable signal-to-interference ratio (dotted) for a hypothetical budget. The SIR for the narrowband source is below the threshold for modes $ k \gtrsim 0.9 \text{ }h\text{ Mpc}^{-1}$. This means that modes beyond this value are excluded by a 1 mJy narrowband source in a budget drawn from this EoR model. In other words, a budget of a single narrowband source of integration flux density equal to 1 mJy is sufficient if this EoR model is accurate, 10\% errors are considered acceptable, and desired measurements are on modes $k \lesssim 0.9 \text{ }h\text{ Mpc}^{-1}$. On the other hand, the SIR for the DTV source is greater than 100 for most modes shown, indicating that a budget of a single 1 mJy two-channel DTV source in the final power spectrum integration is sufficient for the modes shown if only 1\% errors are acceptable and the EoR model is accurate. Alternatively, a 3 mJy DTV source will be acceptable if the error budget is 10\%, but this error budget will be closer to saturation on more modes than in the narrowband case. We relate the quantities shown in this figure to what can be expected from ensembles of RFI sources and sources that appear intermittently in \S\ref{sec:ens}, highlighting the connection between snapshot and integration flux density of sources.}
    \label{fig:lim_comp}
\end{figure*}

\subsubsection{Designing a Budget for a Single Source}

One could parameterize a budget for a single RFI source of given emission profile, a given EoR model, and certain modes by defining a signal-to-interference ratio (SIR) as a function of $k$:
\begin{equation}
    \text{SIR}(k) = \frac{\Delta^2_\text{EoR}(k)}{\Delta^2_\text{RFI}(k)} ,
\end{equation}
and demanding that this ratio be greater than a prescribed value. The power spectrum of a single RFI source is proportional to the square of its apparent flux, but will scale differently as a function of $k$ depending on the frequency dependence of the source, meaning that the final budget for a source will be a function of $k$. For a narrowband source, this relation is given exactly by
\begin{equation}
    \Delta^2_\text{narrow}(k) = \zeta(f_0)(I_0^\text{app})^2\frac{k^3}{2\pi^2}
\end{equation}
where 
\begin{equation}
    \zeta(f_0) = \frac{1}{V_{\mathcal{M}}}\Psi(f_0)^2\kappa(f_0)^2r_\parallel(f_0)^4\beta^2.
\end{equation}
We did not find an exact analytic formula for a DTV power spectrum in this work, particularly after applying a Blackman-Harris window. However, examining Figures \ref{fig:freq_to_power_BH} and \ref{fig:1d_sim} closely, one can see that the envelope for the DTV power spectrum goes roughly as $k^{-2}$,\footnote{This is nearly exact in the theoretical unwindowed case, where a suitable approximation using the sinc function and its derivative describes the Fourier transform of the DTV brightness, implying the leading order term in the power spectrum goes as $k^{-2}$.} i.e.
\begin{equation}
    P_{DTV}(k) \propto \frac{(I_0^\text{app})^2}{k^2}
\end{equation}
indicating that, to reasonable approximation,
\begin{equation}
    \Delta^2_\text{DTV}(k) \propto (I_0^\text{app})^2k.
\end{equation}
These theoretical figures yield two facts: (1) the SIR scales inverse-quadratically with the brightness of the RFI source in any case, and (2) a lower SIR can generally be expected at higher $k$. 

While these theoretical scaling relations are conceptually helpful, power spectrum measurements involve complicated pipelines that can alter the RFI power spectrum compared to these theoretical predictions. As shown in \S\ref{sec:simulation}, there is not a strong discrepancy between the theoretical RFI power spectra and their simulation counterparts. However, for the purposes of drawing an accurate budget, it is better to compare the expected EoR signal to injected RFI signals processed by a 21-cm power spectrum pipeline. To this end, Figure \ref{fig:lim_comp}(a) shows simulated power spectra for two different 1 mJy RFI sources alongside the fiducial EoR model in this work. The simulation pipeline emulates the analysis used in \citet{Barry2019b}, so we show similar modes as were used in that work. Figure \ref{fig:lim_comp}(b) shows the SIR for these two different cases as a function of $k$, along with a flat line at a hypothetical SIR threshold of 10, exemplifying a hypothetical maximum error tolerance of 10\%. Due to the different scaling of the RFI power spectra, one will arrive at different budgets depending on which types of RFI sources contaminate the measurement. For example, assuming errors greater than 10\% of the fiducial EoR model are unacceptable, a 1 mJy narrowband source would be acceptable for modes $k \lesssim 0.9 \text{ }h\text{ Mpc}^{-1}$, and a two-channel DTV source would pose no problems for the modes shown. On the other hand, if only 1\% errors or less are considered acceptable, a narrowband source would exclude $k \lesssim 0.5 \text{ }h\text{ Mpc}^{-1}$, and a 1 mJy two-channel DTV source would nearly saturate this error tolerance on all modes shown. 


\subsubsection{Including Integration and Ensemble Effects}
\label{sec:ens}

An observed RFI source may not be present in every snapshot used within a power spectrum integration. Since snapshots are averaged together in an integration, the apparent flux of a source may be diluted in the full integration relative to the original contaminated snapshots. Additionally, even if a source appears in a very consistent location relative to the telescope, over enough time this will not be a consistent location in celestial co-ordinates. In this way an RFI source may be smeared over the sky and fail to perfectly cohere with itself. This means that the per-snapshot fluxes of RFI sources can in many cases be higher than the mJy level and yet still resemble Figure \ref{fig:lim_comp} as long as there is sufficient dilution. For a highly consistent source, the dilution factor is equal to the fraction of contaminated snapshots in the integration. Any realistic source is unlikely to be consistently located in celestial co-ordinates over the course of a single season, but may reappear at old locations if multiple seasons are combined. Sources that appear in different locations due to seasonal variation or some other reason may instead be considered as many individual sources each diluted by a factor equal to the number of snapshots, which can then be analysed under the dynamics of a source ensemble. Summarizing, when integrating many snapshots together, one may dilute the apparent flux by a factor, $\chi$, that depends on the consistency of the source relative to the observing strategy:
\begin{equation}
    I_0^\text{app, integration} = \chi I_0^\text{app, snapshot}
    \label{eq:chi_dil}
\end{equation}

Another important consideration when developing an RFI budget is the nature of ensembles. Given the occupancy study in \citet{Wilensky2019}, it is relatively likely that any given snapshot contains some extremely faint RFI of some variety. Therefore, a power spectrum integration consisting of many averaged snapshots probably contains an ensemble of RFI emitters. The results in Appendix \ref{sec:app} allow us to quantitatively relate the single-source and ensemble power spectra. We repeat some important results here. 

First, for modes measurable by a physical instrument, the expected power of a random ensemble narrow RFI point-sources uniformly distributed over the sky is the incoherent sum of their individual powers. This can be phrased in terms of the average squared flux of the sources:
\begin{equation}
    \langle P_\text{ensemble}\rangle \propto N\langle(I_s^\text{app})^2\rangle
    \label{eq:inc_prop}
\end{equation}
$N$ is the number of emitters in the ensemble. Without knowing the exact apparent flux distribution and number of emitters, we cannot exactly know the average allowable apparent flux for an incoherent ensemble. However, a many-source ensemble is likely to probe this expectation to within some statistical fluctuation, and so we can infer possible allowances from the above relation. On the other hand, a very particular distribution of RFI emitters can add power coherently for particular modes in the power spectrum. Perfectly coherent and constructive interference is summarized with the relation
\begin{equation}
     P_\text{ensemble} \propto (\sum_{s=1}^NI_s^\text{app})^2 = (I_\text{total}^\text{app})^2.
     \label{eq:coh_sum}
\end{equation}
We note that this is equivalent to the power of a single source whose apparent flux is equal to the total apparent flux of the ensemble. Of course, for certain modes and combinations of sources, destructive interference is possible as well. Combining this relation with Figure \ref{fig:lim_comp}, we see that if the total RFI flux density after dilution can be kept at roughly the mJy level or lower, the EoR model is accurate, and there are not other systematic effects, EoR detection will certainly be feasible even in the worst case scenario of total constructive interference. 

\begin{table*}
    \centering
    \caption{EoR RFI budgets under different circumstances. For narrowband sources, these give 10\% or less fractional excess power relative to the fiducial EoR model for $k \lesssim 0.9$ $h$ Mpc$^{-1}$. For two-channel DTV sources, these give 1\% or less fractional excess power on modes considered in this work. All occupancies and flux densities are expressed per individual RFI source. The integration flux density refers to the flux density of a source in the final integrated spectrum, while the snapshot flux density reflects how bright the source was in its original snapshot. Note that for the final row, 100 incoherent copies of the source appear in the final integration.}
    \label{tab:budgets}
    \begin{tabular}{c|c|c|c|c|c}
        \hline
         Number of Sources & Number of Snapshots & Coherence & Occupancy & Snapshot Flux Density & Integration Flux Density \\
         \hline
         311 & 1029 & Coherent & 0.1\% & 3.3 mJy & 3.2 $\mu$Jy \\
         311 & 1029 & Incoherent & 0.1\% & 58 mJy & 57 $\mu$Jy \\
         1 & 10000 & Coherent & 0.1\% & 1 Jy & 1 mJy \\
         1 & 10000 & Incoherent & 1\% & 1 Jy & 100 $\mu$Jy \\
         \hline
    \end{tabular}
\end{table*}

With these effects in mind, we develop RFI budgets for some hypothetical ensembles, given the fiducial model shown in Figure \ref{fig:lim_comp}. We design the budgets so that they resemble what is shown in Figure \ref{fig:lim_comp}. Since RFI flagging is commonly performed on a per-snapshot basis, a per-snapshot budget may be more readily applicable than a full integration budget. To this end, we describe RFI budgets for a given ensemble in terms of an allowed snapshot flux density per source and an allowed integration flux density per source. The example RFI budgets and parameters that determine them are summarized in Table \ref{tab:budgets}.

As a simple example, consider a single source that appears in a consistent celestial location in 10 snapshots out of a 10000-snapshot integration (0.1\% of the data). Since the source appears in a consistent celestial location in every snapshot, there is a single source in the final integrated image. To yield the same budget as indicated illustrated in Figure \ref{fig:lim_comp}, we demand that the SIR be the same as the single mJy RFI source. This reduces to equation \ref{eq:chi_dil}, with $\chi=0.001$. Therefore, its power spectrum will resemble Figure \ref{fig:lim_comp} if its per-snapshot flux density is 1 Jy. Alternatively, suppose this source appears in uniformly distributed locations on the sky at the same apparent brightness for the sake of simplicity, but occupies ten times as many snapshots. The final integrated image then contains 100 incoherent copies of the RFI source, each diluted by the number of snapshots ($N=100$, $\chi=0.0001$). Invoking equation \ref{eq:inc_prop}, which does not need ensemble brackets when the source fluxes are all identical, we arrive at the relation
\begin{equation}
    P_\text{ensemble} \propto N\frac{(I_0^\text{app, snapshot})^2}{N_\text{snapshot}^2}.
    \label{eq:inc_copy}
\end{equation}
This works out to be exactly the same power as a single 1 mJy RFI source, thus yielding the same SIR. In this alternative example, the RFI source is afforded more occupancy within the budget since its spatial inconsistency makes it less problematic in the final integration. Summarizing, a single bright, consistently missed RFI source can be sustainable so long as it is sufficiently diluted by the observing strategy.

Now consider the data set in the \citet{Barry2019b} limit. Out of the original 1029 snapshots proposed for the integration, 311 snapshots were found to contain residual DTV interference. Assuming one source per contaminated snapshot, we analyse two example realizations of this ensemble below.  

For the first example, let us assume this ensemble sums power incoherently. Following similar logic as in the preceding paragraphs (applying equation \ref{eq:inc_copy} and demanding an identical SIR), we find that if each emitter in this ensemble has a snapshot flux density as low as 58 mJy, corresponding to an integration flux density of 57 $\mu$Jy, the ensemble SIR will resemble Figure \ref{fig:lim_comp}(b). We note that despite the ensemble power spectrum resembling that of a single 1 mJy source, the total RFI integration flux density in this incoherent ensemble is substantially higher at 17.6 mJy. This demonstrates how incoherent ensembles can allow more total integrated RFI flux density for the same SIR.

Alternatively, consider a situation where sources appear in consistent locations on the sky, such as reflections from regularly scheduled aircraft flying due North-South as in \citet{Wilensky2019}. In this particular example, the sources will exhibit coherence for East-West modes. If East-West baselines are highly favored in the analysis, then there will be strong constructive interference in the power spectrum. In the worst case scenario of perfect constructive interference, the total integration flux density budget of 1 mJy can be split over the 311 coherent sources, as in equation \ref{eq:coh_sum}.\footnote{This is mathematically equivalent to considering it as a single source with dilution equal to 311/1029, similar to the very first example ensemble in the preceding paragraphs.} This ensemble of sources will resemble Figure \ref{fig:lim_comp}(b) if the snapshot flux densities of each source are as low as 3.3 mJy, corresponding to an integration flux density of 3.2 $\mu$Jy per source.\footnote{Given that the effect of RFI removal was noticeable at the $10^4$ mK$^2$ level, the brightest sources found by \textsc{SSINS} in this data set must have been substantially brighter than these hypothetical figures.} In these examples, the total number of contaminated snapshots is quite high, and so the per-source allowable RFI brightness is quite low. Given the relative remoteness of the MWA, we expect comparable overall occupancy or worse in most radio telescopes. This low tolerance emphasizes the need to effectively filter RFI and mitigate its effects within EoR data sets.

The exact dilution factor is strongly dependent on the RFI environment, length of integration, and telescope operations (pointing schedule, etc.). Without knowing the exact source of RFI and its mechanism of arrival, there is a great deal of uncertainty in this dilution factor. Setting an accurate per-snapshot budget requires a detailed study of the RFI environment of an EoR telescope. Imaging of contaminated snapshots to understand RFI propagation mechanisms can help determine the consistency of different RFI emitters. This can help inform modifications to telescope operations and data cuts in order to prevent observation of consistent RFI in long integrations.

\section{Conclusions}
\label{sec:conc}

We used theoretical calculations and end-to-end simulations to investigate the level to which RFI occludes EoR detection. For simulation, we used the \textsc{FHD}/$\varepsilon$\textsc{ppsilon} pipeline, with the MWA as our simulated instrument. There was a strong correspondence between the theoretical contamination estimate and the simulated results, verifying the conceptual framework used in this analysis. We conclude that relatively low levels of RFI contamination are sufficient to overwhelm the EoR signal in the 21-cm power spectrum.

Specifically, simulations show that a single narrowband source of 1 mJy apparent flux density in the final power spectrum integration offers excess $\Delta^2$ that is cubic in $k$, with greater than 10\% fractional errors relative to an optimistic EoR model for $k \gtrsim 0.9 \text{ }h\text{ Mpc}^{-1}$. The contamination of this single source scales quadratically as its flux density, and so a single narrowband source of flux density 10 mJy can begin to overwhelm the EoR signal on modes $k \gtrsim 0.3 \text{ }h\text{ Mpc}^{-1}$. This flux density reflects the brightness of the source after forming a power spectrum integration, rather than its brightness in a single snapshot. Depending on the observing strategy and consistency of the source, this may be a strong dilution relative to its snapshot flux density. However, for long power spectrum integrations, it is unlikely that only a single source will be present in the final integration, and so the average allowed apparent flux density of any one RFI source in the final integration may be substantially lower than this number depending on how many sources are actually present in the measurement set.  Averaged over realizations, a random angularly uniform ensemble of narrowband sources is expected to add power incoherently. Furthermore, certain ensembles may add power coherently for certain modes. The power spectrum of a single source of a given flux density, e.g. 1 mJy, represents the maximum power spectrum of a coherent ensemble whose total flux density is the given flux density of that single source. Ultimately, this means that if the assumed optimistic EoR model used here is accurate, a total RFI apparent flux density budget of $\lesssim$ 1 mJy in the final power spectrum integration will be sufficient for attempting to measure modes $k \lesssim 0.9 \text{ }h\text{ Mpc}^{-1}$ assuming a 1 - 10\% error budget depending on the type of RFI that dominates the ensemble. 

This work assumes a foreground avoidance strategy, and so does not focus on lower-order modes where the EoR signal may be stronger. If foregrounds can be successfully mitigated on these modes, it is possible that a less stringent RFI budget can be deduced. On the other hand, the EoR model used is relatively optimistic, and a more pessimistic EoR model will produce a stricter budget.

The size of contamination relative to the faintness of the offending RFI source helps explain the noticeable general improvement of the 21-cm power spectra made by removing RFI-contaminated observations in \citet{Barry2019b}. Furthermore, we expect the results of this analysis to be fairly generic between experiments, particularly predictions for narrowband RFI contamination and any contamination levels within the EoR window. This strongly motivates quantifying the effectiveness of RFI excision implementations, increasing their sensitivity if necessary, and even possibly implementing post-excision mitigation strategies. 

\section*{Acknowledgements}
 We thank the anonymous referee for providing very helpful comments during review. We also thank the development teams of \textsc{pyuvdata}, \textsc{numpy}, \textsc{scipy}, \textsc{matplotlib}, \textsc{FHD}, and $\varepsilon$\textsc{ppsilon} which enabled this work. This work was directly supported by NSF grants AST-1643011, AST-1613855, OAC-1835421, and AST-1506024. NB is supported by the Australian Research Council Centre of Excellence for All Sky Astrophysics in 3 Dimensions (ASTRO 3D), through project number CE170100013. This scientific work makes use of the Murchison Radio-astronomy Observatory, operated by CSIRO. We acknowledge the Wajarri Yamatji people as the traditional owners of the Observatory site. Support for the operation of the MWA is provided by the Australian Government (NCRIS), under a contract to Curtin University administered by Astronomy Australia Limited. 
 
 \section*{Data Availability}
 No new data were generated or analysed in support of this research.

\bibliographystyle{mnras}
\bibliography{main}

\appendix
\section{Narrowband Point Source Ensembles}
\label{sec:app}
The relative ease of power spectrum calculation for narrowband point sources allows one to derive some useful facts. The first is that the power spectrum of any ensemble is bounded from above by the coherent addition of their powers. The second is that a random ensemble of narrowband sources, uniformly distributed over the sky, is expected to sum powers incoherently. The incoherent sum of powers serves as an expected lower bound on the power of the ensemble. The power spectrum of an ensemble of $N$ such sources is given by
\begin{equation}
    \begin{split}
        P_\text{ensemble} = \frac{1}{V_{\mathcal{M}}}\Psi(f_0)^2\kappa(f_0)^2r_\parallel(f_0)^4\beta^2\\
        \times\sum_{s=1}^N\sum_{s'=1}^NI_s^{\text{app}}I_{s'}^{\text{app}}\text{e}^{-\text{i}r_\parallel(f_0)\bold{k_\perp}\cdot(\boldsymbol{\theta_s} - \boldsymbol{\theta_{s'}})}.
        \label{eq:p_ens}
    \end{split}
\end{equation}
The double summation can be split into "diagonal" terms, where $s = s'$, and "cross" terms, where $s \neq s'$. Taking advantage of the symmetry in the indices, we can write this splitting in the following way:
\begin{equation}
    \begin{aligned}
        &P_\text{ensemble} = \frac{1}{V_{\mathcal{M}}}\Psi(f_0)^2\kappa(f_0)^2r_\parallel(f_0)^4\beta^2\\
        &\times\bigg(\sum_{s=1}^N[I_s^{\text{app}}]^2 + 2\sum_{s=1}^N\sum_{s' > s}^N[I_s^{\text{app}}I_{s'}^{\text{app}}\cos(r_\parallel(f_0)\bold{k_\perp}\cdot(\boldsymbol{\theta_s} - \boldsymbol{\theta_{s'}}))]\bigg).
    \end{aligned}
\end{equation}
Since cosine is less than or equal to 1 everywhere, the term above in large parentheses satisfies the inequality
\begin{equation}
    \begin{aligned}
        &\sum_{s=1}^N[I_s^{\text{app}}]^2 + 2\sum_{s=1}^N\sum_{s' > s}^N[I_s^{\text{app}}I_{s'}^{\text{app}}\cos(r_\parallel(f_0)\bold{k_\perp}\cdot(\boldsymbol{\theta_s} - \boldsymbol{\theta_{s'}}))]\\
        \leq &\sum_{s=1}^N[I_s^{\text{app}}]^2 + 2\sum_{s=1}^N\sum_{s' > s}^N[I_s^{\text{app}}I_{s'}^{\text{app}}] = I_0^2,
    \end{aligned}
    \label{eq:NB_ineq}
\end{equation}
where
\begin{equation}
    I_0 = \sum_{s=1}^NI_s^{\text{app}}
\end{equation}
is the total apparent flux of the ensemble. This upper bound is a case of total constructive interference of all the sources in the ensemble. Depending on the source distribution and the wave mode of interest, the cross terms may interfere in a number of ways. For instance, two sources displaced from one another in the direction that is perpendicular to a given spatial mode will necessarily interfere totally constructively for that spatial mode, which can be seen by the dot product between $\bold{k_\perp}$ and $\boldsymbol{\theta_s}-\boldsymbol{\theta_{s'}}$ in the argument of the cosine. On the other hand, there exist other combinations of sources that exhibit total destructive interference for that same mode. The net effect for a power spectrum measurement depends crucially on the exact baseline distribution of the interferometer and the nature of the RFI environment. For instance, in \citet{Wilensky2019}, multiple instances of reflective aircraft flying North to South over the MWA were found, which can be thought of as an ensemble of sources with North-South displacements. These sources tend to produce stronger measurements on East-West baselines. On the other hand, a substantially less remote site might experience a distribution of RFI emitters that is more or less uniform over locations on the sky. Despite not resembling the physical circumstances of certain instruments and RFI environments, we calculate the expected power of a random ensemble of narrowband point-sources that are distributed uniformly and independently over a small patch of sky in order to gain intuition about the nature of possible coherence of RFI sources. We will also consider the true flux distribution as being independent of the location of the source on the sky.

Assuming the primary beam of the instrument to have some characteristic opening angle, $\theta_H$, we can consider sources distributed over a small square patch of the sky of side length $2\theta_H$. Ignoring curvature of the sky, the probability density of a source can then be written
\begin{equation}
    f(\boldsymbol{\theta} | \theta_H) = \frac{1}{(2\theta_H)^2}\Pi\bigg(\frac{\theta_x}{2\theta_H}\bigg)\Pi\bigg(\frac{\theta_y}{2\theta_H}\bigg).
\end{equation}
Using the assumption of independent placement, we can write the joint probability distribution of two sources, labeled $s$ and $s'$ as the product of the above equation for the different sources:
\begin{equation}
    f(\boldsymbol{\theta_s}, \boldsymbol{\theta_{s'}} | \theta_H, \theta_H) = f(\boldsymbol{\theta_s} | \theta_H)f(\boldsymbol{\theta_{s'}} | \theta_H).
\end{equation}
Recalling that $I_s^{\text{app}}=A(\boldsymbol{\theta_s})I_s$, we integrate over this probability distribution and find that the expected power of the ensemble is given by
\begin{equation}
    \langle P_{\text{ensemble}}\rangle = \zeta(f_0)\bigg(N\langle \tilde{A}(\boldsymbol{\theta})^2\rangle\langle I^2\rangle + N(N-1)\langle I\rangle^2\frac{|\tilde{A}(\bold{u})|^2}{(2\theta_H)^4}\bigg),
    \label{eq:exp_power_ens}
\end{equation}
where $\tilde{A}(\boldsymbol{\theta})$ is notation meant to indicate the beam clipped at the opening angle, $\tilde{A}(\bold{u})$ is its Fourier transform as a function of baseline separation vector, and $\zeta(f_0)$ is just the prefactor outside the sums in equation \ref{eq:p_ens}:
\begin{equation}
    \zeta(f_0) = \frac{1}{V_{\mathcal{M}}}\Psi(f_0)^2\kappa(f_0)^2r_\parallel(f_0)^4\beta^2.
\end{equation}
We have switched from $\bold{k_\perp}$ to $\bold{u}$ since quantities involving the primary beam are more readily understood in this frame.

We define the expected coherency factor, as a function of baseline, as
\begin{equation}
    c(\bold{u}) = \frac{|\tilde{A}(\bold{u})|^2}{(2\theta_H)^4}.
\end{equation}
For a non-negative beam on the sky that is peak-normalized to 1, one can show that this coherence function is bounded between 0 and 1, with its maximum attained at the origin in the $uv$-plane. As an example, for a beam that is equally sensitive out to the opening angle (a top-hat beam), this function takes the form
\begin{equation}
    c(\bold{u}) = \text{sinc}^2(2\pi\theta_Hu)\text{sinc}^2(2\pi\theta_Hv).
\end{equation}
This is equal to 1 at the origin and falls off as a power law in any given direction. As a result, baselines longer than the inverse width of the top-hat primary beam tend to experience dramatically less coherence than shorter baselines. We expect this intuition to transfer to the case of a primary beam with more realistic angular dependence by way of typical Fourier transform reasoning. What we see, then, is that uniformly distributed sources are expected to add incoherently for most baselines used in analysis, rather than conveniently adding destructively, or much worse, constructively. Moreover, we can easily scale the calculated or simulated power spectrum of a single source such as in \S\ref{sec:results} by a linear factor if we instead want to consider an ensemble of emitters.

Note that in equation \ref{eq:exp_power_ens}, other than the assumption of independence of angular position, we have made no assumptions about the nature of the flux distribution of the ensemble. In principle, one could inform this distribution using studies such as \citet{Offringa2013} and \citet{Sokolowski2016}. This would allow one to specifically relate the average flux squared (incoherent term) to the average flux of the ensemble, and, in turn set an average allowed apparent flux of individual RFI sources within an integration.

\end{document}